\documentclass{nature}

\newcites{latex}{Supplementary Reference}

\title{A Solid State Source of Photon Triplets Based on Quantum Dot Molecules}

\author{Milad Khoshnegar$^{1,2,3}$, Tobias Huber$^{4}$, Ana Predojevi{\'c}$^{4}$, Dan Dalacu$^{5}$, Maximilian Prilm{\"u}ller$^{4}$, Jean Lapointe$^{5}$, Xiaohua Wu$^{5}$, Philippe Tamarat$^{6}$, Brahim Lounis$^{6}$, Philip Poole$^{5}$, Gregor Weihs$^{2,4}$, Hamed Majedi$^{1,3}$}

\begin{document}

\maketitle

\begin{affiliations}
 \item Department of Electrical and Computer Engineering, University of Waterloo, Waterloo, Ontario N2L 3G1, Canada
 \item Institute for Quantum Computing, University of Waterloo, Waterloo, Ontario N2L 3G1, Canada 
 \item Waterloo Institute for Nanotechnology, University of Waterloo, Waterloo, Ontario N2L 3G1, Canada
 \item Institut f{\"u}r Experimentalphysik, Universit{\"a}t Innsbruck, Technikerstr. 25, 6020 Innsbruck, Austria
 \item National Research Council of Canada, 1200 Montreal Road, Ottawa, Ontario K1A 0R6, Canada
 \item Universit{\'e} Bordeaux, LP2N Institut d'Optique and CNRS, Talence F-33405, France

\end{affiliations}

\begin{abstract}

Producing advanced quantum states of light is a priority in quantum information technologies. In this context, experimental realization of multipartite photon states would enable improved tests of the foundations of quantum mechanics as well as implementations of complex quantum optical networks and protocols. It is favourable to directly generate these states using solid state systems, for simpler handling and the promise of reversible transfer of quantum information between stationary and flying qubits. Here we use the ground states of two optically active coupled quantum dots to directly produce photon triplets. The formation of a triexciton in these ground states leads to a triple cascade recombination and sequential emission of three photons with strong correlations. We record 65.62 photon triplets per minute under continuous-wave pumping, surpassing rates of earlier reported sources. Our structure and data paves the way towards implementing multipartite photon entanglement
and multi-qubit readout schemes in solid state devices.
\end{abstract}

\section*{Introduction}

With the rise of quantum technologies, generalized quantum key distribution (QKD) protocols\cite{Lo2014,Diamanti2016,Gobby2004} based on multipartite entangled states could be stepping stones towards realizing real-world quantum networks.\cite{Gisin2007} While remarkable progress has been made on creating single photons and entangled photon pairs, multipartite correlated photon states are usually produced in purely optical systems by postselection techniques or cascading, with extremely low efficiency and exponentially poor scaling.\cite{Rauschenbeutel2000,Pan2001,Zhao2004} The most widespread technique for generating multipartite photon correlations relies on spontaneous parametric down conversion (SPDC) with low conversion efficiency\cite{Hubel2010,Hamel2014,Bouwmeester1999,Guerreiro2014} and restricted scalability, which limits its production rate and applications. Moreover, in order to generate multipartite correlated photons, most schemes based on SPDC use the interference of photon pairs created by independent Poissonian sources and post-select the favoured subset of output photon states,\cite{Wang2016,Bouwmeester1999,Eibl2003,Zhao2004,Yao2012} which significantly adds to the probabilistic nature of the process and the uncorrelated background light. 

In contrast, quantum dots offer the most practical route in building scalable quantum architectures and their efficiency reaches almost unity per excitation pulse, enabling high count rates. The ground state of a single quantum dot hosts at most two bright excitons,\cite{Moreau2001} a biexciton, which can be controlled coherently\cite{Jayakumar2014} to produce correlated photon pairs in a so-called cascade recombination process. Thus the creation of multipartite photon correlations in a single quantum dot requires exploiting energetically higher shells and phonon-mediated processes under heavy pumping, which lead to inevitable dephasing, line broadening and poor photon correlation visibility.\cite{Persson2004,Schmidgall2014} The coupled s shells of a quantum dot molecule (QDM), however, render additional excitonic states suitable for increasing the number of correlated photons possibly using coherent schemes. The wavefunctions of photogenerated excitons localized in the QDM s shells are coupled via molecular hybridization and Coulomb interactions,\cite{Bayer2001} thus the radiative recombination of such molecular excitons will naturally prepare correlated photons. The hybridization of carrier wavefunctions in a QDM is a strong function of the interplay between dot composition and interdot spacing. Nanowire-embedded quantum dots offer controllable size and composition,\cite{Dalacu2012} which enable engineering of the QDM interdot coupling and its spectral properties. In addition, the core-shell structure alleviates the propagation and extraction of the optical modes that carry  photons\cite{Reimer2012} and promises more efficient detection of the photons emitted from higher-order excitons, which is a requisite in our photon correlation measurements.\cite{Michler2000} 

In the following, we demonstrate the creation of photon triplets using a QDM positioned inside an epitaxially-grown photonic nanowire. The photoluminescence spectrum of  our QDM shows two sets of resonances calibrated by the QDM material and size. We identify these resonances by conducting a series of power-dependent and time-resolved spectroscopy experiments along with magneto-photoluminescence and photon correlation measurements. We observe a clear bunching-antibunching pattern when the photon correlations between each pair of triexciton, biexciton and exciton resonances are measured, which implies the emission of a photon triplet through a triple cascade recombination process. Employing the molecular s shells of QDM aids us achieve a far better photon correlation visibility than previous attempts in single quantum dots.\cite{Persson2004,Schmidgall2014} The photon triplet emission rate is estimated by conducting triple coincidence experiments in both continuous-wave and pulsed excitation regimes, showing a remarkable improvement compared to the direct creation of triplets in SPDC-based schemes. A realization of photon triplets from a triexciton forming in a QDM serves as an elementary step for the direct generation of multiphoton entanglement, which has been so far limited to photon pairs in solid state systems.\cite{Stevenson2006}

\section*{Results}

\noindent\textbf{Quantum dot molecule structure}

\begin{figure}
\centering
\includegraphics[scale=0.2]{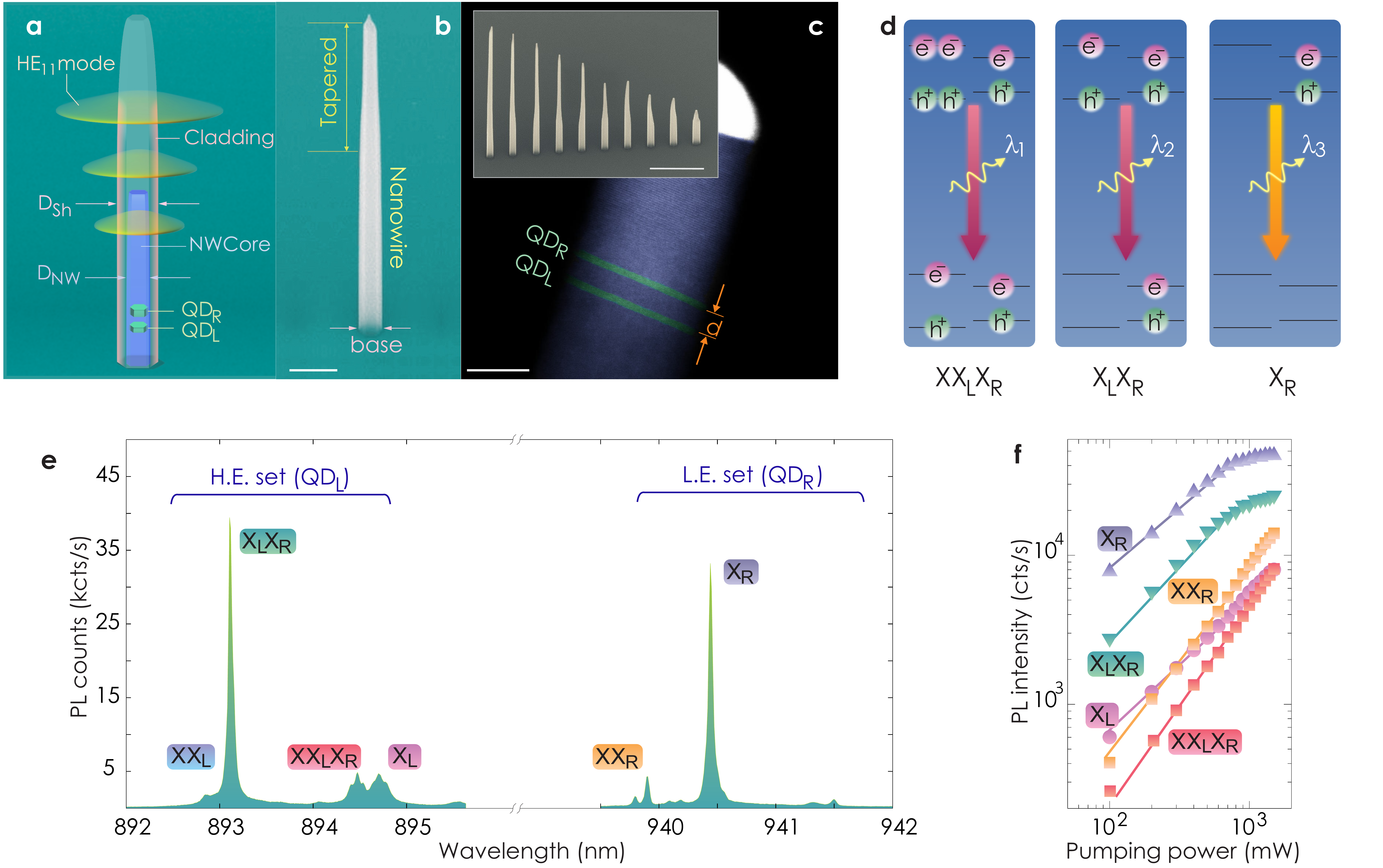}\caption{\textbf{Structure and spectrum of Nanowire-QDM} a) Schematic of quantum dot molecule (QDM) embedded inside a clad nanowire. The best suited nanowires consist of a thin core region $D_\mathrm{NW}=18$-$20~ \mathrm{nm}$ surrounded by a thick InP cladding (shell) $D_\mathrm{Sh}=250~\mathrm{nm}$ that waveguides at least one principal optical mode at QDM emission wavelengths $\sim$894 and $\sim$940 nm. (b) False-coloured scanning electron microscopy image of a spatially isolated nanowire with hexagonal cross-section incorporating a single QDM. The scale bar is 500 nm. (c) False-coloured transmission electron microscopy image of an InP nanowire (core) grown on (111)B substrate in wurtzite phase embedding two In(As)P quantum dots separated by $\sim$7 nm. The scale bar is 10 nm. Inset: The nanowires are site-controlled allowing for excellent isolation of QDM spectrum from inhomogeneous broadening. The scale bar is $1~\mu\mathrm{m}$ (d) Triple sequential transitions: carrier configurations of high-energy (H.E.) triexciton, separated biexciton and low-energy (L.E.) exciton. (e) Optical spectrum of QDM comprising two prominent features at $\sim$894 nm and $\sim$940 nm. Inset: photoluminescence intensity of the QDM resonances showing linear or superlinear dependence on the pump power.\label{fig:Structure-Spectrum}}
\end{figure}

Our QDM is composed of two InAs$_x$P$_{1-x}$ segments ($x\approx0.15$ and $0.25$) embedded inside an InP photonic nanowire that incorporates core and cladding regions\cite{Huber2014} (Fig. \ref{fig:Structure-Spectrum}a). The thick cladding of 250 nm in diameter aids funnelling the QDM emission into the fundamental $\mathrm{HE_{11}}$ mode\cite{Bleuse2011} to be guided out toward the collection optics. The cladding is gently tapered ($2^\circ$) at its apex to improve the photon extraction efficiency (Fig. \ref{fig:Structure-Spectrum}b). The molecule contains two ${h_\mathrm{D}}\sim$2.5-3 nm thick and ${D_\mathrm{NW}}\sim$18 nm in diameter dots as confirmed by transmission electron microscopy (TEM) imaging (Fig. \ref{fig:Structure-Spectrum}c). The growth of the second dot $\mathrm{QD_R}$ is seemingly influenced by the strain field caused by the formation of the first dot $\mathrm{QD_L}$ during the molecular beam epitaxy (MBE) process, giving rise to some  compositional asymmetry of the molecule. Notice that even though the hybridization energy itself can exceed several tens of meV in strongly-coupled double dots,\cite{Bayer2001} an important part of the $s$-shell splitting in the molecule studied here is induced by the above material composition change. Such an inherent asymmetry aids the localization of the heavy hole wavefunctions mainly inside the two individual dots rather than evenly spread throughout the molecule.\cite{Stinaff2006} The proximity of dot and barrier compositions however leads to a comparatively weaker localization of the electron, and its orbital partially diffuses into the neighbouring dot. An interdot spacing of $\sim$8-10 nm was initially targeted in the vapour-liquid-solid (VLS) growth mode, however the Arsenic tailing in our dots possibly reduces the effective separation $d$ down to $\sim7$ nm. Considering the low Arsenic concentration ($0.15<x<0.25$) of the dot segments, a thinner spacing would lift the barrier and aid delocalization of electrons, or would promote the directional nonresonant tunneling in the QDM,\cite{Gawarecki2010} whereas a larger spacing would impair the electron hybridization and interdot coupling. The yield of finding a suitable QDM in our investigated samples was 10$\%$.

\noindent\textbf{Spectroscopy measurements and interdot coupling}

In our experiment, the formation of a triexciton in the QDM entails the photongeneration of a biexciton ($X\!X$) in one quantum dot ($\mathrm{QD_{L}}$) along with an exciton ($X$) in the neighbouring dot ($\mathrm{QD_{R}}$) under continuous optical pumping. The predominant coupling mechanism among the two dots can be explained either via the wavefunction hybridization and Coulomb interactions,\cite{Bayer2001,Bester2004} or the direct energy transfer of excitons (F{\"o}rster process),\cite{Govorov2003} or  nonresonant phonon-assisted tunneling. The direct transfer of excitons is caused by long-range Coulomb interactions and typically occurs if the interdot energy splitting is small, at most a few meV. As shown later, the energy detuning of the constituting quantum dots is several tens of meV in our molecule because of its structural asymmetry, hence the direct exciton transfer has a negligible impact on the interdot coupling here.\cite{Rozbicki2008,Govorov2003} Moreover, the nonresonant tunneling of carriers in QDMs is a function of the phonon spectral density, thus depends on the wavefunction overlap and particularly the energy difference of the states involved in the transition. This implies that any carrier tunneling between the two detuned $s$ shells of our QDM would require multiple acoustic phonon processes.\cite{Stavrou2006} We will later demonstrate that nonresonant tunneling plays a minor role in the interdot communication here, and therefore wavefunction correlation is the primary source of coupling.

The studied QDM shows two distinguished high energy (H.E.) and low energy (L.E.) sets of spectral resonances at $\sim$894 nm and $\sim$940 nm corresponding to its molecular $s$-shell direct transitions (Fig. \ref{fig:Structure-Spectrum}e). The formation probability of optically active indirect excitons should be small owing to the molecule asymmetry and rather single-dot-confined holes.\cite{Beirne2006} In addition to the conventional exciton ($X\!_{\mathrm{L}}$ or $X\!_{\mathrm{R}}$) and biexciton ($X\!X\!_{\mathrm{L}}$ or $X\!X\!_{\mathrm{R}}$) direct transitions belonging to $\mathrm{QD_{L}}$ and $\mathrm{QD_{R}}$, there exist energy-shifted biexciton and exciton transitions, $X\!X\!_\mathrm{L}\!X\!_\mathrm{R}$ at $\lambda_1=894.5$ nm and $X\!_\mathrm{L}\!X\!_\mathrm{R}$ at $\lambda_2=893.1$ nm emerging due to Coulomb interaction with $X\!_{\mathrm{R}}$ at $\lambda_3=940.9$ nm. The carrier configuration related to the transitions creating the photon triplet are shown in Fig. \ref{fig:Structure-Spectrum}d. They are assigned by acknowledging that bright interdot recombination is unlikely and that the $X\!X\!_\mathrm{L}\!X\!_\mathrm{R}$ and $X\!_\mathrm{L}\!X\!_\mathrm{R}$ resonances are located in the H.E. set. For simplicity, we name these two latter transitions triexciton and separated biexciton, respectively. The power-dependent PL intensity of the above resonances exhibit the expected linear and superlinear regimes for both series of regular and energy-shifted excitons and biexcitons, respectively (Fig. \ref{fig:Structure-Spectrum}f). The emergence of $X\!_{\mathrm{R}}$ at the lowest excitation levels makes the conditional formation of separated biexciton and triexciton in $\mathrm{QD_L}$ more likely than that of $X\!_{\mathrm{L}}$ and $X\!X\!_{\mathrm{L}}$. $X\!X\!_{\mathrm{L}}$ grows on the shoulder of the neighbouring $X\!_\mathrm{L}\!X\!_\mathrm{R}$ resonance at higher excitation levels, which hinders resolving its power dependence over the entire range.

In order to understand the possible effect of nonresonant carrier tunneling, we performed a time-resolved micro-PL experiment on the present QDM and another double dot, $\mathrm{DD}_2$, with identical single dot specifications, but an increased interdot spacing of over 30 nm in $\mathrm{DD}_2$ to eliminate the coupling. The lifetime of the single exciton $X\!_{\mathrm{L}}$ of the QDM was measured $\sim2.8\pm0.2$ ns (similar value can be inferred by comparing $X\!_\mathrm{L}\!X\!_\mathrm{R}$ and $X\!_\mathrm{R}$ lifetimes as shown in Supplementary Note 4), whereas the same $X\!_{\mathrm{L}}$ resonance of $\mathrm{DD}_2$ lasted $\sim2.5\pm0.2$ ns. In general, the decay time $\tau_\mathrm{d}$ of the exciton $X\!_\mathrm{L}$ in a molecule, where nonresonant tunneling from $\mathrm{QD_L}$ to $\mathrm{QD_R}$ continuously takes place, is given by $1/\tau_\mathrm{d}=1/\tau_\mathrm{r}+1/\tau_\mathrm{t}$, where $\tau_\mathrm{r}$ is the exciton radiative lifetime and $1/\tau_\mathrm{t}$ is the tunneling rate. The fact that $\tau_\mathrm{d}$ ($=2.8\pm0.2$ ns) and $\tau_\mathrm{r}$ ($=2.5\pm0.2$ ns) are comparable within the accuracy of our experiment suggests that the impact of nonresonant tunneling between the $s$ shells of our QDM is negligible and perhaps a reverse mechanism exists between the $s$ shell of $\mathrm{QD_L}$ and the nearby $d$ shells of $\mathrm{QD_R}$ appearing at slightly higher energies in the spectrum. The nonresonant electron tunneling is however enhanced at a small enough spacing ($d<3$ nm), where the barrier is lifted and $\mathrm{QD_L}$ is steadily emptied showing weak PL intensity (see Supplementary Note 2). The above observations indicate that the coupling in our QDM forms primarily via the hybridization of electron wavefunctions. 

\noindent\textbf{Triple coincidence experiments} 

\begin{figure}
\centering
\includegraphics[scale=0.26]{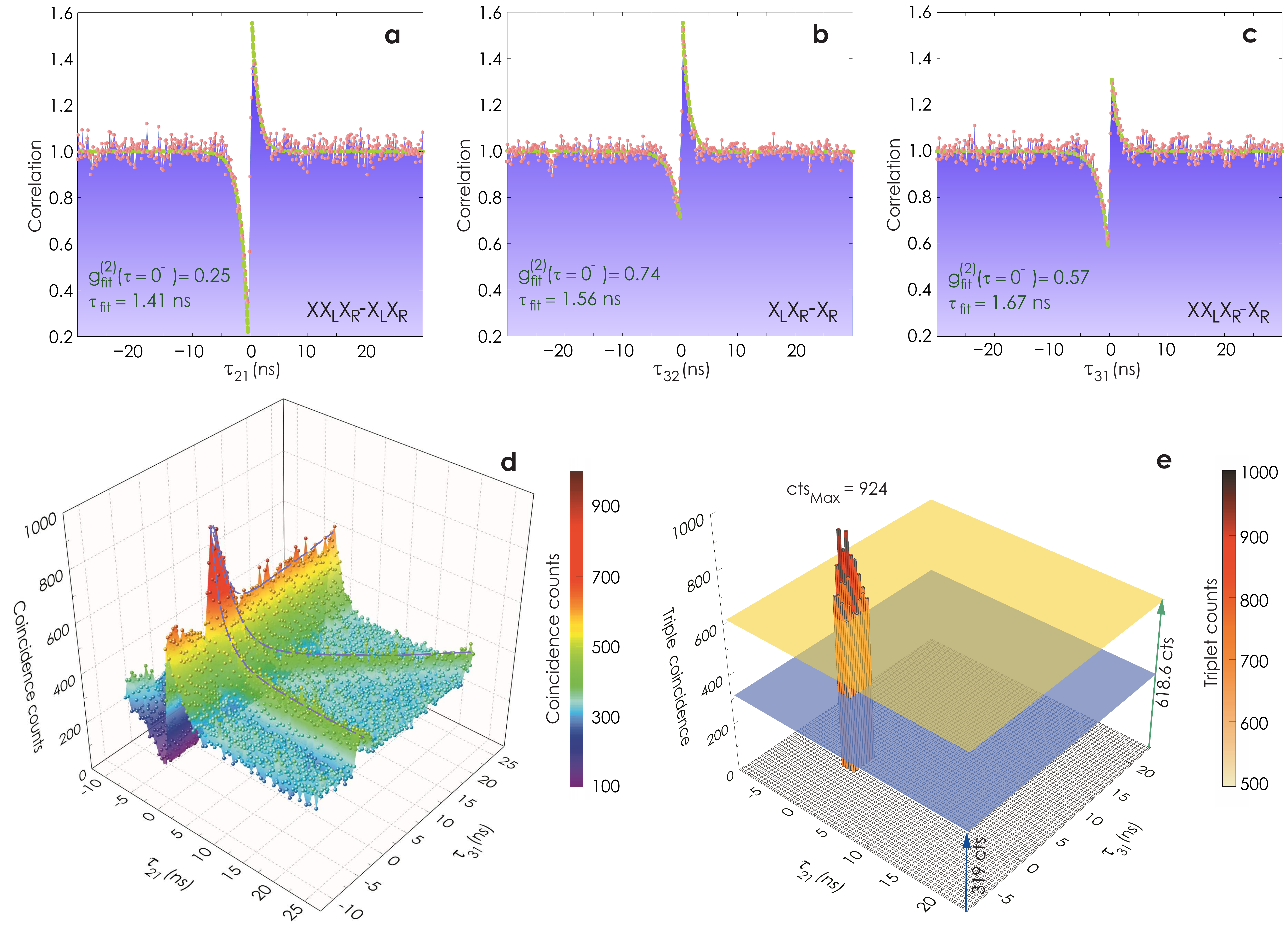}\caption{\textbf{Dual-channel cross correlations and triple coincidence histogram} (a-c) Normalized cross correlations of $X\!_\mathrm{R}$, $X\!_\mathrm{L}\!X\!_\mathrm{R}$ and $X\!X\!_\mathrm{L}\!X\!_\mathrm{R}$ versus delay time measured at the high excitation power density ($\mathrm{PD}$) of $6.9~\mathrm{W/mm^2}$ showing a sequential triple cascade recombination. The antibunching dips are fitted with $g^{(2)}_\mathrm{fit}(\tau)=1-ae^{(\tau/\tau_\mathrm{fit})}$ ($\tau<0$), where the anticorrelation floor is limited by the background noise. (d) The triple coincidence histogram (total recording time 3~h) was measured at $\mathrm{PD}=460~\mathrm{mW/mm^2}$ and is plotted versus $\tau_{21}$ and $\tau_{31}$, linearly interpolated with a color-mapped surface. The threefold coincidence peak near the origin signifies the strong temporal correlations of the emitted photons. e) Events above the two-fold cascade threshold from d) without interpolation. The threshold level (yellow plane) was determined as the (peak) value of $g_{21}^{(2)}(0^+)$ averaged over $t_\mathrm{D3}$ outside the triple coincidence window. For comparison the expected level of accidental triplet events is shown in blue.\label{fig:CC}}
\end{figure}

The true character of $X\!X\!_\mathrm{L}\!X\!_\mathrm{R}$, $X\!_\mathrm{L}\!X\!_\mathrm{R}$ and $X\!_\mathrm{R}$ were fully identified by conducting a series of magneto-photoluminescence measurements to confirm that the Zeeman splitting of their spin fine structure comply with the theoretically predicted values, as presented in Supplementary Note 3. The next step was to measure the second-order autocorrelation function\cite{Santori2002} of each individual resonance and the cross-correlation functions\cite{Moreau2001} $g^{(2)}_{\alpha\beta}(\tau)$ between various pairings ($\alpha$,$\beta$) of distinct resonances ($\tau_{\alpha\beta}=t_{D\alpha}-t_{D\beta}$ denotes the delay time between photon detections). These correlations can provide insight into the coupling strength and the nature of the lines.\cite{Gerardot2005} An autocorrelation experiment on every QDM resonance in our sample verified their low multiphoton emission probability by featuring an antibunching dip at $\tau=0$. Among all possible cross correlations of QDM resonances, the ones measured between $X\!X\!_\mathrm{L}\!X\!_\mathrm{R}$, $X\!_\mathrm{L}\!X\!_\mathrm{R}$, and $X\!_{\mathrm{R}}$ are of highest interest for the photon triplet characterization. A triplet state comprising temporally correlated photons $\ket{\lambda_1\lambda_2\lambda_3}$ originates from a triple sequential cascade in the QDM (Fig. \ref{fig:Structure-Spectrum}d). In our correlation setup illustrated in the Supplementary Note 1, a diffraction grating separates the $X\!X\!_\mathrm{L}\!X\!_\mathrm{R}$, $X\!_\mathrm{L}\!X\!_\mathrm{R}$, and $X\!_{\mathrm{R}}$ photons towards the detectors D1, D2 and D3. All cross-correlations $g^{(2)}_{\alpha\beta}(\tau)$ of the above three resonances feature an asymmetric bunching-antibunching behaviour\cite{Moreau2001} as expected for cascade transitions (Figs. \ref{fig:CC}a-c). The cross-correlations between $X\!X\!_\mathrm{L}\!X\!_\mathrm{R}$ (or $X\!_\mathrm{L}\!X\!_\mathrm{R}$) and $X\!_\mathrm{R}$  are fitted with $g^{(2)}_\mathrm{fit}(\tau)=1-ae^{(\tau/\tau_\mathrm{fit})}$ ($\tau<0$) resulting in $g^{(2)}_\mathrm{fit}(0^{-})=$ 0.71 (0.59) considerably smaller than unity, which indicates that the system is indeed a molecule rather than two separate dots. Here, the non-zero level of correlation at $\tau = 0^-$ can be explained by, first, the temporal dynamics of the transitions,\cite{Gerardot2005} which depends on the ratio between their pumping rate $\mathrm{W_p}$ and decay rates $\Gamma_X$ as further scrutinized in the Supplementary Note 4 (increasing $\mathrm{W_p}/\Gamma_X$ lifts the antibunching floor and suppresses the bunching peak); second, the parasitic background caused by the phonon sideband of the neighbouring weak spectral lines or stacking fault states. The effect of such background noise is more pronounced once photons from the L.E. set contribute to the correlations, because they are collected by a fiber with a twice larger core that also collects more background emission (see Methods). Similarly, the cross correlations among the remaining L.E. and H.E. resonances featured above antibunching characteristic, in contrast to cross correlations in DD2 which showed no sign of antibunching.

In order to prove that the QDM actually emits a photon triplet, we conducted a triple coincidence experiment\cite{LahmamBennani1989} by sending detector pulses D1 (as Start), D2 (as Stop1) and D3 (as Stop 2) into a time-tagging device. The time-resolved histogram versus $\tau_{21}$ and $\tau_{31}$ features a fully random contribution due to uncorrelated photons (319 counts) along with three contributions each coming from two correlated photons and the third being accidental resulting in a total level of (partially correlated) triple events of 618.6 counts. We observe a large number of threefold coincidences in the vicinity of zero time delay (Fig. \ref{fig:CC}d) above the partially correlated events. We record 20744 photon counts in total (including 8932 random background counts) integrated in 3 hours in the coincidence window of  $\tau_{21}\in\{-0.768,1.28\}~\mathrm{ns}$ and $\tau_{31}\in\{-1.28,2.304\}~\mathrm{ns}$ (see Fig. \ref{fig:CC}e). To ensure that the photon triplet generation rate is not overestimated, we subtract all the random or partially correlated events, which leaves us with 11812 photon triplets corresponding to an average detection rate of 65.62 triplets per minute. We estimate that only $0.023\%$ of all photon triplets could be detected because of the low detection efficiency of our detectors, $\eta_\mathrm{D}=\eta_\mathrm{D1}\;\eta_\mathrm{D2}\;\eta_\mathrm{D3}$ ($\eta_{\mathrm{D1}}=25\%$, $\eta_{\mathrm{D2}}=25\%$ and $\eta_{\mathrm{D3}}=15\%$ at the respective wavelengths), along with non-ideal extraction efficiency $\eta_{\mathrm{C}}=46\%$ (see Methods), fibre coupling efficiency ($\eta_{\mathrm{F}}=85\%$) and grating efficiency ($\eta_{\mathrm{G}}=75\%$). The above photon triplet rate is, to the best of our knowledge, the highest recorded rate exceeding the values reported for direct generation of photon triplets via cascaded SPDC under cw pumping.\cite{Hubel2010,Guerreiro2014}

\begin{figure}
\centering
\includegraphics[scale=0.27]{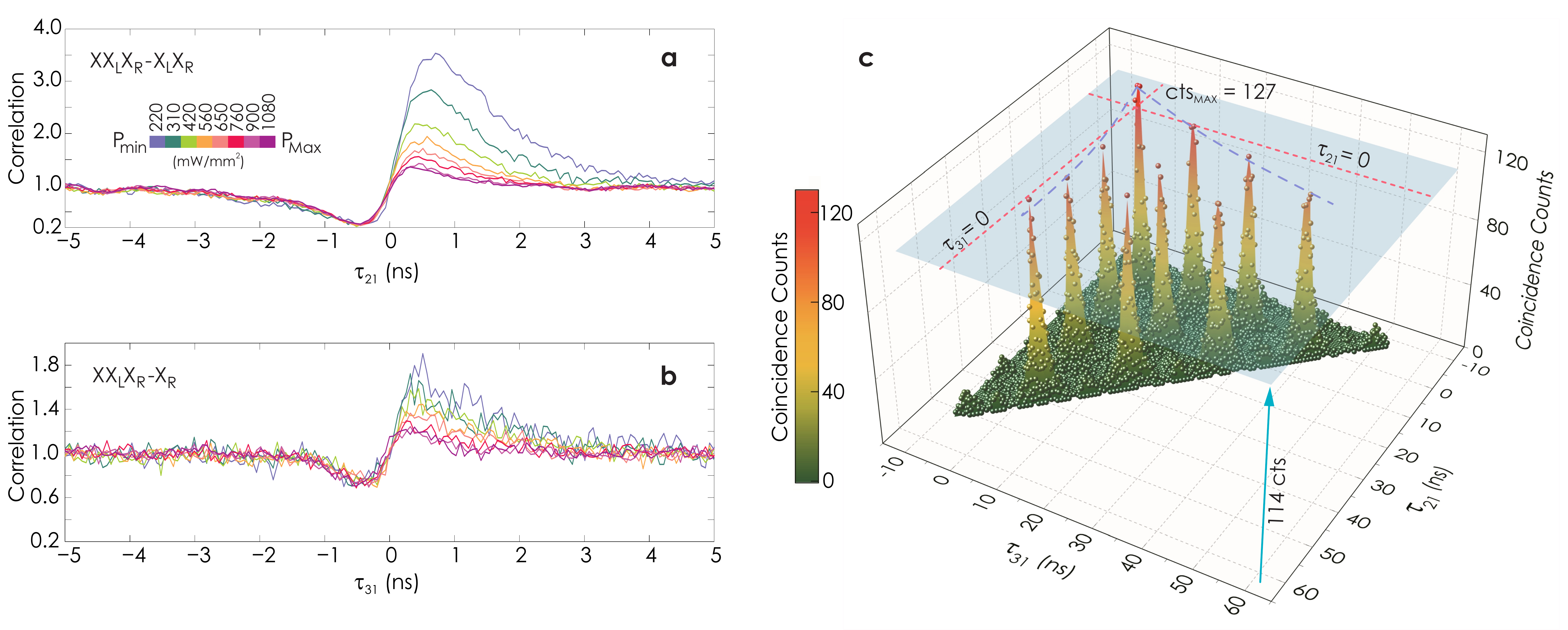}\caption{\textbf{Bunching visibility and triple coincidence counts under pulsed excitation} (a-b) Normalized cross correlations of $X\!X\!_\mathrm{L}\!X\!_\mathrm{R}$ resonance with $X\!_\mathrm{L}\!X\!_\mathrm{R}$ and $X\!_\mathrm{R}$ resonances measured at eight increasing power densities starting from 220~$\mathrm{mW/mm^2}$. The histograms are color-coded according to the applied pumping levels. (c) The triple coincidence histogram measured in 80 mins plotted versus $\tau_{21}$ and $\tau_{31}$ and linearly interpolated with a color-mapped surface. The blue plane, at 114 counts, indicates the threshold level separating genuine photon triplet counts from the partially correlated photon counts.
\label{fig:Powerseries}}
\end{figure}

In general, the bunching peak $g_{\alpha\beta}^{(2)}(0^+)$ of a cascade decreases with the excitation rate, because the ratio of true cascade events versus individual excitations becomes less favorable, as has been observed in regular biexciton-exciton cascades of single quantum dots.\cite{Kuroda2009} We examined this behavior by applying increasing levels of pump power while recording the cross correlations between the triexciton and the other two resonances (see Figa. \ref{fig:Powerseries}a and b). The measurements were conducted in a regime where the resonance PL intensity to background hardly changed, thus the variation in the bunching peak was mainly a function of the ratio between the excitation rate $\mathrm{W_p}$ and transition lifetimes ($1/\Gamma_X$). The difference between the bunching visibility of $X\!X\!_\mathrm{L}\!X\!_\mathrm{R}\text{-}X\!_\mathrm{L}\!X\!_\mathrm{R}$ and $X\!X\!_\mathrm{L}\!X\!_\mathrm{R}\text{-}X\!_\mathrm{R}$ cross-correlations in Figs. \ref{fig:Powerseries}a and b also originates from the inequality of this ratio, $\mathrm{W_p}/\Gamma_{X}$, in $X\!_\mathrm{L}\!X\!_\mathrm{R}$ and $\!X\!_\mathrm{R}$ resonances, together with their unequal PL intensity measured by the silicon avalanche photodiodes at different wavelengths (see Methods). The suppression of bunching visibility with increasing the excitation power agrees with the results of our theoretical model based on the time propagation matrix method,\cite{Gerardot2005} as explained in the Supplementary Note 4, and reconfirms the cascaded nature of the selected transitions.

Finally, we demonstrated the formation of triexciton and creation of photon triplets under the pulsed excitation regime. For this purpose, the QDM was pumped non-resonantly with 2.6 ps pulses at 820 nm in the same cross-correlation setup used for the cw pumping regime (see Methods). Figure \ref{fig:Powerseries}c illustrates the triple coincidence counts versus $\tau_{21}$ and $\tau_{31}$ measured in 80 mins, featuring a central peak located at ($\tau_{21}=0$, $\tau_{31}=0$) and a 2D grid of side peaks with a temporal period of 12.5 ns, equal to the pulse cycle. The coincidence peaks in this histogram have contributions from fully and partially correlated events as previously identified in the cw regime. The central coincidence peak comprises all above contributions along with the fully correlated photon triplets occurring after the first excitation pulse, whereas the side peaks primarily result from the fully accidental and partially correlated events taking place between consecutive pulse excitations. We estimated the maximum number of  partially correlated events at the side peaks to be 114 counts, and thus all the counts above this level and within a 5 ns ($\sim \tau_{X\!X\!_\mathrm{L}\!X\!_\mathrm{R}} + \tau_{X\!_\mathrm{L}\!X\!_\mathrm{R}} + \tau_{X\!_\mathrm{R}}$) time window around the central peak were considered as true photon triplet counts, that is 363 photon triplets in 80 mins (4.53 triplets per min). The lower rate of photon triplet generation here, as compared to the cw regime, could be attributed to the lower average cw equivalent power, which essentially reduces the number of photogenerated carriers in the higher shells that eventually feed the ground state of the molecule within less than 1 ns. Moreover in our method of calculation, the number of detection events considered as a genuine photon triplet count is also a function of the ratio between the pulsed laser repetition rate and $\tau_{X\!X\!_\mathrm{L}\!X\!_\mathrm{R}}$, $\tau_{X\!_\mathrm{L}\!X\!_\mathrm{R}}$ or $\tau_{X\!_\mathrm{R}}$, because longer lifetimes increase the probability of photon correlation between e.g. $X\!_\mathrm{R}$ and re-excited $X\!X\!_\mathrm{L}\!X\!_\mathrm{R}$ or $X\!_\mathrm{L}\!X\!_\mathrm{R}$ from the consecutive cycles. We predict that under the coherent
excitation, the background noise and the amplitude of the side peaks will drastically drop and the maximal triplet count rate will increase up to 17 kHz with the given efficiencies. Nevertheless, the above rate still tops the rates of direct photon triplet generation employing SPDC under pulsed pumping by an order of magnitude.\cite{Krapick2016} 

\section*{Discussion}

Creating entangled photon triplets, as opposed to time correlated ones, remains as the next-step study to our present observations. The prospects of tripartite photon entanglement includes, but are not limited to, multipartite quantum secret sharing and quantum communication protocols,\cite{Gaertner2007,Cleve1999} and third party cryptography. As a relevant example, tripartite time-bin entanglement\cite{Simon2005} can be realized using the spin states of a triexciton bound in a QDM. Time-bin encoding has a clear benefit for long distance quantum communication through optical fibers because the relative phase between each two pulses with a few nanosecond temporal spacing is merely susceptible to a medium varying faster than this timescale. Implementing such kind of entanglement in a QDM, however, demands for the resonant pumping of the triexciton to encode the laser phase onto the emitted photons in a relatively dephasing-free process.\cite{Jayakumar2014} In contrast with the incoherent pulsed excitation, almost a complete elimination of background light is expected under the resonant pumping, and due to the absence of additional intraband relaxation processes the time jitter will be limited to the exciton radiative lifetime. In analogy with single quantum dots, a coherent pulsed excitation of QDM can prepare the triexciton in either of the singlet and triplet spin states, $\mathrm{\ket{0_{XX,L},S_{R}}}$ or $\mathrm{\ket{0_{XX,L},T_{R}}}$, where $\mathrm{\ket{0_{XX}}}=\ket{\uparrow\Downarrow\downarrow\Uparrow}$, $\ket{\mathrm{S}}=1/\sqrt(2)(\ket{\downarrow\Uparrow}-\ket{\uparrow\Downarrow})$ and $\ket{\mathrm{T}}=1/\sqrt(2)(\ket{\downarrow\Uparrow}+\ket{\uparrow\Downarrow})$, and $\{\uparrow,\downarrow\}$  ($\{\Uparrow,\Downarrow\}$) denote the electron (heavy hole) spin localized in the left (L) or the right (R) quantum dot. $\mathrm{\ket{0_{XX,L},S_{R}}}$ ($\mathrm{\ket{0_{XX,L},T_{R}}}$ could then decay to $\{\mathrm{(S_{L},S_{R})},\mathrm{(T_{L},S_{R})}\}$ ($\{\mathrm{(S_{L},T_{R})},\mathrm{(T_{L},T_{R})}\}$) biexciton states, followed by a second and eventually a third transition to $\mathrm{\ket{0_{L},S_{R}}}$ ($\mathrm{\ket{0_{L},T_{R}}}$) and the ground state $\mathrm{\ket{0_{L},0_{R}}}$ (see Supplementary Note 6 for the detailed diagram). These transition paths provide four sets of triple decays emitting three polarized photons $\ket{\mathrm{H_1H_2H_3}}$, $\ket{\mathrm{V_1V_2H_3}}$, $\ket{\mathrm{H_1H_2V_3}}$ and $\ket{\mathrm{V_1V_2V_3}}$ in H and V linear basis, either of which could be utilized, for example, to create Greenberger-Horne-Zeilinger\cite{Bouwmeester1999} (GHZ) time-bin entangled photon triplets. Here, the coherent pumping of triexciton is feasible through either employing three different colored lasers in resonance with the transitions of interest or pumping virtual levels.\cite{Jayakumar2014} In either case, the output pulse of the lasers is split into two pulses, early ($e$) and late ($l$). At sufficiently low pumping powers, a triexciton is formed by either early or late pulses and the wavefunction of the three emitted photons can be represented as $1/\sqrt(2)(\ket{e_{1}e_{2}e_{3}}+e^{i\phi}\ket{l_{1}l_{2}l_{3}})$, where $\phi$ is the sum of the phases in the pumping interferometers and, along with the phase of the triplets, it should be later analysed by three output interferometers. A tomography experiment is then conducted to resolve the three-photon density matrix by recording the coincidences between the output photons of the analysing interferometers. 

At the first glance, our rather low emission rate of photon triplets under the incoherent pulsed excitation regime might imply an  inefficient generation of entangled photon triplets using QDMs in future. However, we predict a drastic improvement of  the photon triplet counts under the resonant excitation due to the profound suppression of the background noise and accidental coincidences. In this case, the triplet generation rate is approximately given by $\eta_\mathrm{ex}\;\eta_\mathrm{D1}\;\eta_\mathrm{D2}\;\eta_\mathrm{D3}\;\eta_\mathrm{C}^3\;\eta_\mathrm{G}^3\;\eta_\mathrm{F}^3\;n_\mathrm{P}$, where $n_\mathrm{P}$ denotes the pulse repetition rate, and $\eta_\mathrm{ex}$ is the excitation probability of the triexciton, which can potentially reach up to $90\%$ with an optimized pulse length as previously demonstrated for the biexciton.\cite{Huber2016} Under such circumstances, improving the detection efficiency, for example by employing near-ideal superconducting nanowire photodetectors,\cite{Marsili2013} or enhancing the light extraction efficiency, by embedding a reflective layer under the nanowire base,\cite{Reimer2012} could potentially boost the integrated triplet counts by two orders of magnitude during the state tomography measurements. 

In conclusion, we have demonstrated that a triexciton bound in a QDM can originate time-ordered photon triplets in a cascaded process. We expect to improve the triplet generation rate by reducing the inter-dot energy splitting, deterministic coherent pulsed excitation of the triexciton to reduce the background, enhanced collection efficiency and the use of more efficient detectors. Triple excitons forming in the $s$ shells of a QDM should, in priciple, benefit from far better coherence properties than the $p$-shell exciton in single quantum dots, because their coherence time $T_2$ is not subject to the dephasing caused by the $p$-to-$s$ phonon scattering relaxation. The necessity of populating higher shells in single quantum dots also requires strong optical pumping, which further adds to the spectral diffusion and the photon decoherence problem. The inhomogeneous broadening observed in our current QDM samples, however, arise from the stacking faults in the nanowire, which function as efficient charge traps and cause the spectral wandering.\cite{Reimer2016} The density of such stacking faults are expected to be reduced by improving the MBE growth conditions at higher temperatures (500 $^\circ$C) in the near future. With the earlier  demonstration of quantum-dot-based quantum key distribution (QKD),\cite{Waks2002} our device facilitates the implementation of multiparty quantum secret sharing on integrated semiconductor chips.

\section*{Methods}
\label{Met}

\textbf{Nanowire-QDM Fabrication}\\
The InP nanowires with embedded In(As)P quantum dots are grown using selective-area VLS epitaxy. The nanowires are grown on a SiO$_2$-patterned (111)B InP substrate. The pattern consists of circular holes defined using electron-beam lithography and hydrofluoric acid wet-etch. A single gold particle is deposited in each hole using a self-aligned lift-off process, with the size of the particle determined by the hole size and the thickness of deposited gold. We employ chemical beam epitaxy (CBE) with trimethylindium (TMI) and pre-cracked PH$_3$ and AsH$_3$ sources. The growth temperature is $420^\circ$C. Two growth modes are utilized to grow a nanowire core, which defines the quantum dots, and a shell, which defines the cladding of photonic nanowire. The nanowire core is grown under a reduced PH$_3$ flow resulting in an untapered InP nanowire with a diameter corresponding to the gold catalyst particle, approximately 20\,nm in this work. The nanowires are pure phase wurtzite with less than 1 stacking fault per micron.\cite{Dalacu2012} The double In(As)P quantum dots are grown by switching the group V species from phosphorous to arsenic to grow the first dot, switching back to phosphorous to grow the InP spacer, then switching back to arsenic to grow the second dot while maintaining a constant flux of TMI. Samples were grown with quantum dot growth times of 2.5 and 3 seconds, and with spacer times of 10, 15, and 60 seconds. The interdot spacing for a given growth time between dots depends on the core diameter due to a diameter-dependent growth rate.\cite{Dalacu2009} By using a diameter-dependent growth model\cite{Dalacu2009} we calculate an interdot separation of 8-20 nm for core diameters of 18-24 nm. Details of the spacer-dependent interdot coupling are beyond the scope of this work and will be published elsewhere. The spacer of QDM studied here is 10 seconds ($\sim$ 7-8 nm) that provides the optimum coupling. The nanowire shell is grown by increasing the PH$_3$ flow rate by a factor of three, which reduces the Indium adatom migration length and promotes deposition on the nanowire sidewall facets. The shell is grown to reach base diameters of 250\,nm, resulting in nanowires with heights of $\sim$ 5 $\mu$m and tapers of $\sim 2^\circ$.

\noindent\textbf{Optical Experiments}\\
The sample is cooled down to 6 K in a customized and thermally stabilized liquid-Helium continuous-flow cryostat. The QDM is nonresonantly excited either by a cw or a ps-pulsed Ti:Sapphire laser at 820 nm with 8.4 ps (or 2.6 ps for the cross-correlation measurement) pulse duration (80 MHz repetition rate) slightly above the wurtzite InP band gap 1.49 eV (832 nm) and the donor-acceptor recombination level 1.44 eV (861 nm). We excite the QDM via a separate objective rather than the collecting objective even though this is not reflected in the setup schematic in Supplementary Note 1. The molecule luminescence is collected using an objective lens with numerical aperture (NA) of 0.7 and dispersed by grating monochromators with spectral resolution of $\sim$0.01 nm to split the spectral lines and send the respective photons into separate avalanche photodiodes (APD). APDs are identical with $\sim$300 ps temporal resolution and $\sim$25$\%$ ($\sim$15$\%$) detection efficiency at 893 nm (940 nm). The combination of spectrometer and charge coupled device (CCD) camera enables performing in-situ spectroscopy during the recording of counts in the correlation measurement setup (composed of APDs and ps time-tagging module). Only two APDs register the photon counts to conduct the autocorrelation and the conventional dual-channel cross correlation analysis, whereas all the three APDs are in use for the triple coincidence experiment. In the dual-channel correlation measurements, the H.E. set resonances are cross correlated utilizing two 5 micron core optical fibres for photon collection. In the triple coincidence experiment, we collected from the L.E. set using a single mode fibre with 9 micron core optimized for the telecommunication wavelength, that operates as a multimode fibre at 940 nm. The multimode character improves the collection efficiency without the requirement for an optimized mode matching. However, owing to the small core radius the background light picked up from $X_\mathrm{R}$ is suppressed and the antibunching dip in triple coincidence histogram is improved compared to a 125 $\mu$m core multimode fiber. To estimate the extraction efficiency of nanowire, we calculate the probability of biexciton-exciton coincidence $\eta_{\mathrm{D1}}\eta_{\mathrm{D2}}\eta_{\mathrm{C}}^2\eta_{\mathrm{G}}^2\eta_{\mathrm{F}}^2$ from the dual-channel cross-correlation histogram to be 0.54$\%$, which results $\eta_{\mathrm{C}}$ to be $46\%$. To produce the power-dependent cross correlation histograms in the cw excitation mode, we started from 220$~\mathrm{mW/mm^2}$ (with 4~$\mu \mathrm{m}$ spot size and excitation objective tilted 22$^{\circ}$ from the optical table axis) and raised the pump power to linearly increase the shifted biexciton $X\!X\!_\mathrm{L}\!X\!_\mathrm{R}$. Therefore the pump power scales up approximately in a quadratic fashion until the $X\!_\mathrm{L}\!X\!_\mathrm{R}$ resonance is saturated. In order to resolve the associated lifetimes, the QDM is heavily pumped within each pulse using the Ti:Sapphire laser in a way that its resulting spectrum exactly resembles the one under cw excitation. In the triple coincidence experiment under the pulsed excitation, the pumping power was adjusted to 40 $\mu$W (with 2.6 ps pulse duration) which translates to a power density of 905 mW/mm$^2$ per pulse. The temporal resolution of the detectors D1, D2 and D3 was set to 512 ps. The laser spot size was approximately 7.5 $\mu$m on the sample (measured perpendicular to the beam, which was aligned under an angle of approximately 50$^\circ$ with respect to the nanowire axis). For the magneto-optical measurements the setup remains unchanged except that the cryostat is replaced by a continuous flow exchange gas cryostat with a 7 T split-pair superconducting magnet. The QDM was excited co-linear to the collection through the collection objective with a Ti:Sapphire laser. For the mixed Voigt-Faraday (tilted) configuration the sample was rotated $12^{\circ}$ inside the cryostat.

\section*{Data Availability Statement}

The data that support the findings of this study are available from the corresponding author upon request.

\section*{Author Contribution}
M.K. conceived the idea and developed the theory. M.K. and D.D. designed the QDM. D.D., J.L., X.W. and P.P. fabricated the nanowire-QDMs. M.K., A.P and T.H. designed the photon statistics experiments. T.H., A.P. and M.P. accomplished the photon correlation measurements. M.K., A.P., and B.L. designed the magneto-optical measurements conducted by A.P., P.T., and M.P. T.H., A.P. and M.P. compiled the data and M.K. carried out the data analysis and wrote the manuscript with feedback from all co-authors. G.W. and H.M. supervised the project and contributed to the manuscript.

\section*{Acknowledgement}
This work was funded by NSERC Discovery Grant Program, National Research Council Canada and the European Research Council. M.K. thanks NSERC for partial support through the CryptoWorks21 fellowship. T.H. thanks the Austrian Academy of Sciences for receiving a DOC Fellowship. A.P. would like to thank the Austrian Science Fund for the support provided through project number V-375. P.T. and B.L. acknowledge financial support from  the Institut Universitaire de France and the Laphia cluster of excellence (IDEX Bordeaux).

\section*{Competing interests}
The authors declare no competing financial interests.

\section*{Corresponding author}

contact M. Khoshnegar (m3khoshn@uwaterloo.ca)

\bibliographystyle{naturemag.bst}
\bibliography{reference}

\end{document}